# Temperature and field dependence of thermally activated flux flow resistance in $Bi_2Sr_2CaCu_2O_{8+\delta}$ superconductor

Devina Sharma[1,2,*], Ranjan Kumar[2] and V P S awana[1,#]

[1]National Physical Laboratory, Dr. K S Krishnan Marg, New Delhi, India - 110012

[2]Department of Physics, Punjab University, Chandigarh, India – 160014

## Abstract

We study the temperature dependence of the resistivity as a function of magnetic field in superconducting transition ($T_c^{onset} - T_c^{R=0}$) region for different $Bi_2Sr_2CaCu_2O_{8+\delta}$ superconducting samples being synthesized using sol-gel method. The superconducting transition temperature ($T_c^{R=0}$) of the studied samples is increased from 32 K to 82K by simply increasing the final sintering temperature with an improved grains morphology. On the other hand, broadening of transition is increased substantially with decrease in sintering temperature; this is because $T_c^{onset}$ is not affected much with grains morphology. Further broadening of the superconducting transition is seen under magnetic field, which is being explained on the basis of thermally activated flux flow (TAFF) below superconducting transition temperature ($T_c$). TAFF activation energy ($U_0$) is calculated using the resistive broadening of samples in the presence of magnetic field. Temperature dependence of TAFF activation energy revealed linear temperature dependence for all the samples. Further, magnetic field dependence is found to obey power law for all the samples and the negative exponent is increased with increase in sintering temperature or the improved grains morphology for different Bi-2212 samples. We believe that the sintering temperature and the ensuing role of grain morphology is yet a key issue to be addressed in case of cuprate superconductors.

*Corresponding author's e-mail: s.sharmadevina@gmail.com

Tel.  +91-11-45609357; Fax.  +91-11-45609310

#For further queries: awana@mail.nplindia.ernet.in (www.freewebs.com/vpsawana)

# 1. Introduction

Although two decades are passed since the discovery [1] of Cu based high temperature superconductors (HTSC), yet there are no major applications based on them. Various superconducting parameters; like high superconducting transition temperature ($T_c$), high upper critical field ($H_{c2}$) and sustainability of critical current density ($J_c$) at elevated magnetic field etc., are the deciding factors for technological applicability of any superconductor. One of the major problems with HTSC is the weak flux pinning and the resultant significant drop in the critical current density with the applied magnetic field [2]. In addition, there is an appearance of resistivity at finite temperatures, well below $T_c$ at extremely small critical current densities, which occurs due to the hopping of thermally active flux bundles across the pinning sites [3].

The flux line dynamics can be divided into three regimes: (i) flux flow; $J > J_c$ (ii) Thermally activated flux flow (TAFF); $J \ll J_c$ and (iii) flux creep; $J \sim J_c$ in the transition region between the two. Flux flow and flux creep are common terms for low temperature superconductors (LTSC) but TAFF has been coined for a new aspect of HTSC. Flux flow phenomena arises, when current in a superconductor is high enough to essentially tear loose the flux lattice and sets it in sliding motion. In HTSC, it has been observed that even for very low currents $J \ll J_c$, the superconductor in the magnetic field displays a linear (ohmic) resistance well below $T_c$. This phenomenon, suggested by Dew-Hughes is known as thermally activated flux flow (TAFF) [4]. TAFF is associated closely with the flux creep model of Anderson and Kim [5] in low current density limit. The resistivity in this regime is given by $\rho = \rho_o(B,T) e^{-U_0/\kappa_B T}$ where, $\rho_o$ is the pre exponential factor, $\kappa_B$ is the Boltzmann's constant and $U_0$ is the characteristic TAFF activation energy, which is slightly temperature and magnetic field dependent. The activation energy show parabolic behavior or exhibits different power law exponent with magnetic field, depending on the type of superconductor or precisely the flux pinning in them [6]. There have been many reports regarding temperature and field dependence of activation energy in various HTSC. Palstra *et al.* [3] reported power law dependence $U_0 \propto H^{-\alpha}$ on field, where as Kucera *et al.* [7] suggested that $U_0 \propto H^{-1/2}(1-T/T_c)$, where $T_c$ is the critical temperature for Bi-2212 thin films. The same relation was suggested by Wagner *et al.* [8] for Bi-2212 thin films. In a recent report by Zhang *et al.* [9], they suggested an empirical form $U_0 \propto H^{-\alpha(H)}(1-t)^{\beta(H)}$.

In the present work, we report the measurement of resistive transition broadening due to thermally activated flux flow in the samples of $Bi_2Sr_2CaCu_2O_{8+\delta}$ superconductor being synthesized by sol gel method and sintered at various temperatures. The superconducting transition width of varying temperature ($840^oC$ to $760^oC$) sintered samples range from 14.6K to 58.0K. It is found that TAFF model well describes the resistive transitions measured at different applied magnetic fields. The field dependence of the activation energy was found to obey different power law given by $U_0(H) \sim H^{-n}$ and linear temperature dependence for variously synthesized samples. Our results flash new light on the role of grain boundaries on TAFF and the resultant superconducting properties of variously synthesized $Bi_2Sr_2CaCu_2O_{8+\delta}$ superconductor.

## 2. Experimental

Samples of $Bi_2Sr_2CaCu_2O_{8+\delta}$ superconductor were synthesized using sol gel method using ethylene-diamene-tetra-acetic acid (EDTA) as a chelating agent [19,20]. High purity of $Bi_2O_3$, $SrCO_3$, $CaCO_3$ and CuO were dissolved in nitric acid to obtain the nitrates of Bi, Sr, Ca and Cu. The obtained solutions of nitrates were mixed and added to an aqueous EDTA. The molar ratio of EDTA to the total metal cation concentration was chosen to be unity. The $p_H$ of the obtained acidic solution was raised to seven by subsequently adding ammonium hydroxide to it. The liquid was stirred and heated continuously at $80^oC$ to result in a transparent viscous gel. On further heating the viscous gel expanded to foam like, and was finally converted into precursor powder. Obtained precursor powder was calcined at $500^oC$ to remove organic impurities. Further sintering of calcined powder at different temperatures i.e. at 760, 780, 820 and $840^oC$ was done to obtain four different samples with different grains morphology.

Phase analysis of the samples was carried out on Rigaku X-ray diffractometer. Microstructural examination of the samples was done using scanning electron microscopy (SEM). The temperature dependence of resistivity was measured by standard four probe method using Physical Property Measurement System (PPMS). Measurements were carried out for the samples in the temperature range of 10 to 110K at different magnetic fields varying from 0 to 14 Tesla.

## 3. Results and discussion

Figure 1 shows the X-Ray Diffraction patterns of the polycrystalline $Bi_2Sr_2CaCu_2O_{8+\delta}$ samples sintered at 780 and 840°C. Though the main phase is Bi-2212, some unreacted small intensity lines are also seen along with minor 2201 phase. In BSCCO systems the intergrowth of various phases viz. Bi-2212, 2201 and 2223 is abundant. In any case the secondary phases are very minor. Also the characteristic low angle (5.8 degree) 2212 peak is seen without intergrowth of Bi-2201 (7.6 Degree) and 2223 (4.8 Degree). This shows that the majority phase formed is Bi-2212 only and is true for all the samples being synthesized at different temperatures between 760 and 840°C. The lattice parameters of the samples are calculated and it was found that though, the *a* parameter remains nearly invariant close to 3.82 Å, the *c* –parameter is increased from 30.5 to 31.0 Å with increase in sintering temperature. We rule out the possibility of changing *c* parameter having any effect on vortex dynamics; because the superconducting onset of all studied samples is nearly same, except for highest temperature synthesized near melt sample, see Table 1.

Figure 2 shows the characteristic SEM pictures of the samples sintered at various temperatures. The magnification of all the SEM images is same. From SEM images, it can be seen that the sample sintered at lowest temperature (760°C) is quite porous with very small sized grains weakly connected to each other. The grains shape is changed to thin flakes like with length and width of few μm when the same sample is sintered at 780 and 820°C. And finally the sample becomes nearly a dense melt when sintered at highest temperature (820°C). It is clear that the grain morphology in terms of better coupling is substantially improved for higher temperature synthesized sample. The message we carry from Figure 2 is that grain growth is improved for higher temperature synthesized samples, which is quite obvious. The sol-gel method has facilitated us to form Bi-2212 phase at as low as 760°C and at higher temperatures of up to 840°C. This way one could vary the grain morphology quite substantially. In fact though the 760°C synthesized sample possesses enormous porous regions, the one synthesized at 840°C is nearly a compact melt.

The study of thermally activated flux flow (TAFF) is manifested as a broadening of the superconducting transitions. In the presence of magnetic field, such broadening is interpreted in terms of energy dissipation caused by vortex motion. The resistance in the TAFF region is caused by the flow of the vortices, which are thermally activated. Thus, resistivity is given by $\rho = \rho_0(B,T)e^{-U_0/\kappa_B T}$ [3], where $U_0$ is the flux flow activation energy, which can be obtained from the slope of the linear part of the ln ($\rho/\rho_0$) versus $T^{-1}$ plot. In the present investigation $\rho_0$ has been taken as normal state resistance at 110K. This choice is physically reasonable as it allows U(T) to go to zero just above the onset of $T_c$, where $\rho_0 = \rho_0(110K)$. $U_0$ is obtained from the limited range of the resistivity data corresponding to TAFF region, where Arrhenius plots of $\rho/\rho_{110K}(T)$ yields a straight line. In order to determine the temperature and field dependence of $U_0$, let us assume that

$$U_0 = c \times a(H) \times b(T) \qquad \ldots\ldots\ldots\ldots(1)$$

Where, a(H) and b(T) are functions of field and temperature respectively and c is a constant. In the present study we will follow the empirical formula given by Zhang et al. [9]

$$U_0 \alpha\ H^{-\alpha(H)} (1-t)^{\beta(H)} \qquad \ldots\ldots\ldots\ldots(2)$$

Where, $t = T/T_x(H)$.

There are previous reports [7,8] in which $T_x(H)$ is reported to be constant (~ $T_c$ at H=0) and independent of magnetic field. But, Palstra *et al.* [3] and Kim *et al.* [10] proposed a magnetic field dependent $T_x(H)$ instead of $T_c$. In the present investigation, $T_x(H)$ is been found by fitting the resistivity data using equations (1) and (2).

Figure 3(a)-(d) shows the resistivity versus temperature plots under magnetic fields of 0, 0.05, 0.1, 0.5, 1, 2, 4, 6, 8, 10, 12 and 14 Tesla for the four sets of $Bi_2Sr_2CaCu_2O_{8+\delta}$ samples, being sintered at 760, 780, 820 and 840°C respectively. Table 1 shows the transition temperature associated with the onset of superconductivity $T_c^{onset}$ of the sample and $T_c^{R=0}$ along with the superconducting transition width ($\Delta T_c$) at no applied magnetic field. It can be seen that superconducting transition is sharper for the higher temperature sintered samples and the same is quite broad for those sintered at low temperatures. The transition width being defined as $T_c^{onset} - T_c^{R=0}$ decreases monotonically from say 58.0K to 14.6K for 760°C and 840°C respectively in zero field. The trend is more or less the same under applied fields as well. The broadening of the resistive transition in a magnetic field for layered superconductors is interpreted in terms of dissipation of energy caused by motion of vortices. So, this substantial broadening of the superconducting transition width is possibly due to depreciation in the vortex flux pinning in lower temperature synthesized samples. Further, in case of Bi-2212@760°C sample, two step superconducting transition can be seen without any applied magnetic field, which could be the manifestation of superconductivity at intra and inter granular level in the sample. But, this two step transition decreases and eventually vanishes for the samples sintered at higher temperatures showing improvement in the grain coupling with sintering temperature.

Since resistivity in the TAFF regime can be written as $\rho=\rho_0(B,T)e^{-U_0/\kappa_B T}$, so the activation energy can be obtained by plotting $\ln(\rho/\rho_o)$ versus $T^{-1}$. Figure 4(a)-(d) shows the Arrhenius plots for the four samples, where activation energy is given by the slope of the linear region of low resistivity. The low resistivity region in the range $10^{-4}$ to $10^{-5}$ $\Omega$ cm$^{-1}$ of each of the plots is fitted linearly to obtain activation energy at various magnetic fields. In order to determine the magnetic field dependence of the TAFF activation energy ($U_0$) for each of samples, the activation energy obtained is plotted as a function of applied magnetic field as shown in Figure 5. It can be seen from the plot that $U_0$ follows power law dependence on magnetic field for all the samples as already reported for the case of HTSC [11-13]. Figure 5 also shows the data fitting of the same curves as per $U_0 = c \times H^{-\alpha(H)}$, using equations (1) and (2). From the curve fitting it is revealed that although $U_0$ exhibit same (power law) behavior with respect to magnetic field, but with slightly different exponents. Table 2 shows the values of c and α obtained from the curve fitting data of Figure 5. The values of both c and α increase monotonically for higher temperature synthesized samples, except that the c value of 840 $^0$C sample is decreased.

The value of α is -0.18, -0.23, -0.41, and -0.45 for samples sintered at 760, 780, 820 and 840°C respectively. It is observed that both the field dependence on TAFF and flux pinning strength increases with increase in the sintering temperatures of the samples. The sample sintered at 760°C exhibits least flux pinning among all the samples and also shows weakest field dependence (α=-0.18). The strongest field dependence is exhibited by sample sintered at 840°C (α=-0.45), suggesting the possibility of collective flux creep dominance [21,22] in it. Also, since α = 0.5 and 1 corresponds to planar and point defect pinning respectively [23], it is possible that flux lines are pinned by collective planar defects in this sample.

Temperature dependence of activation energy is determined by re-plotting the broadening data as $-T(\ln(\rho/\rho_0))$ versus T. Figure 6(a)-(d) shows the temperature dependence of activation energy for the four samples. The activation curves are depicted as non linear function of temperature. This function has an upward kink at temperature $T \leq T_c$ after which $U_0$ attains linear temperature dependence. This region corresponds to the low resistance portion of the Arrhenius curves depicted in Figure 3, i.e. the regime where vortex motion is determined by thermally activated flux hopping. The kink separating the non linear part (low $U_0$) from the linear part (higher $U_o$) of the curve is shown by a hypothetical dashed line in Figure 6. Sudden increase in the activation energy at low temperatures has been reported many times earlier, in particular for HTSC [7,14]. The appearance of the upward kink has been explained by the possibility of showing either the crossover from 2D to 3D vortex nature or transition from flux flow/creep to TAFF region. The linear part above the dashed line was linearly fitted with respect to temperature as per $U_0 \alpha\ a(H) \times (1-t)^{\beta(H)}$ dependence, using equations (1) and (2), where, $t = T/T_x(H)$. It was found that β takes the value unity at all field values for all the samples, hence the temperature dependence was found to be linear for all Bi-2212 samples irrespective of their sintering temperature. Also, $T_x(H)$ is found to be field dependent for each of the sample and is shown in Table 2. The temperature $T_x(H)$ denotes the transition temperature dividing the flux flow and flux creep dissipation regime above and below it respectively. At this temperature, the activation energies become comparable to the thermal energies hence facilitating TAFF.

## 4. Conclusion

The results of the dependence of activation energy ($U_0$) on field and temperature for different temperature synthesized Bi-2212 samples having their superconducting transition widths in range of 14.6K to 58.0K are presented. Though, $U_0$ is linearly dependent on temperature without any applied magnetic field, the same under magnetic field obeys power law dependence with different exponents for all the studied Bi-2212 samples of varying superconducting transition width or grain morphology. Clearly, these results indicate towards the role of the grain morphology (hence, flux pinning) in the TAFF resistive broadening. The TAFF could be controlled to some extent in higher temperature synthesized or better grains coupled Bi-2212 samples. For viable practical high field applications the Bi-2212 HTSC compounds need to be fabricated close to their melting temperature so that the grains connectivity is excellent and the TAFF is controlled to some extent.

## 5. Acknowledgements

Authors thank Prof. R.C. Budhani, Director, NPL and Dr. Hari Kishan, HOD, for their keen interest and encouragement for superconductivity research. Mr. A.K. Sood from the SEM Division of NPL is acknowledged for providing us with the SEM micrographs.

**Figure Captions**

**Figure 1** XRD pattern of Bi-2212 samples sintered at $760^{o}C$ and $840^{o}C$.

**Figure 2** SEM images of the samples sintered at (a) 760 (b) 780 (c) 820 and (b) $840^{o}C$

**Figure 3** Temperature dependence of normalized resistivity in the magnetic field range of 0 - 14 Tesla for samples sintered at (a) 760 (b) 780 (c) 820 and (d) $840^{o}C$.

**Figure 4** Arrhenius plots of the resistive transition of the samples sintered at (a) 760 (b) 780 (c) 820 and (d) $840^{o}C$ in the magnetic field range of 0 - 14 Tesla. Linear part of low resistivity region is fitted to obtain activation energy at various fields.

**Figure 5** $U_0$ dependence on magnetic field for samples sintered at (a) 760 (b) 780 (c) 820 and (d) $840^{o}C$. Dotted lines are the theoretical fit of equation $U_0 = c \times H^{-\alpha(H)}$

**Figure 6** $U_0$ dependence on temperature for samples sintered at (a) 760 (b) 780 (c) 820 and (d) $840^{o}C$ in the magnetic field range of 0 - 14 Tesla. Linear part above dashed line is fitted to determine temperature dependence at various fields.

**Table 1** Lattice parameters and superconducting transition temperatures $T_c^{onset}$, $T_c^{R=0}$ and $\Delta T_c$ of samples sintered at various temperatures

| Sintering Temp (°C) | $a$ (Å) | $c$ (Å) | $T_c^{onset}$ (K) | $T_c^{R=0}$ (K) | $\Delta T_c$ |
|---|---|---|---|---|---|
| 760 | 3.81(2) | 30.51(3) | 89.75 | 31.74 | 58.01 |
| 780 | 3.81(1) | 30.90(2) | 89.80 | 49.78 | 40.02 |
| 820 | 3.82(1) | 31.06(2) | 89.80 | 67.80 | 22.00 |
| 840 | 3.82(3) | 30.74(4) | 96.40 | 81.79 | 14.61 |

**Table 2** Values of parameters, α, β, c and $T_x(H)$ defined in equations (1) and (2)

| | $T_x(H)$ (Kelvin) | | | |
|---|---|---|---|---|
| | Bi-2212@760°C | Bi-2212@780°C | Bi-2212@820°C | Bi-2212@840°C |
| Field (T) | α = 0.19, β = 1 | α = 0.23, β = 1 | α = 0.41, β = 1 | α = 0.45, β = 1 |
| | c = 17.9 | c = 56.5 | c = 80.8 | c = 64.4 |
| 0 | 67.38 | 67.70 | 77.37 | 92.63 |
| 0.05 | 62.18 | 65.71 | 76.64 | 92.01 |
| 0.1 | 57.90 | 66.20 | 74.95 | 91.69 |
| 0.5 | 50.19 | 59.36 | 72.55 | 91.02 |
| 1 | 48.44 | 53.75 | 70.16 | 89.29 |
| 2 | 45.31 | 47.63 | 65.76 | 87.77 |
| 4 | 41.78 | 51.31 | 59.74 | 85.02 |
| 6 | 39.96 | 40.11 | 51.20 | 79.09 |
| 8 | 39.42 | 37.24 | 49.32 | 70.52 |
| 10 | 62.18 | 35.46 | 46.68 | 78.65 |
| 12 | 45.94 | 35.08 | 44.81 | 76.10 |
| 14 | 41.87 | 34.92 | 44.23 | 73.54 |

Figure 1

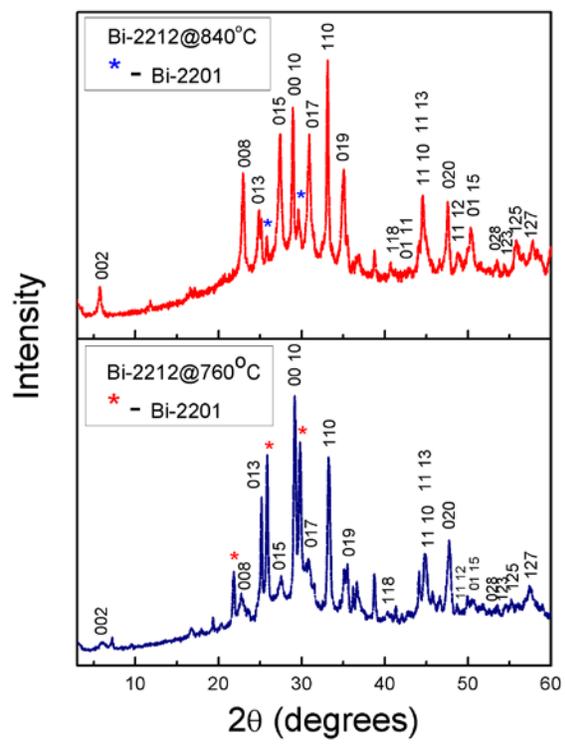

Figure 2a

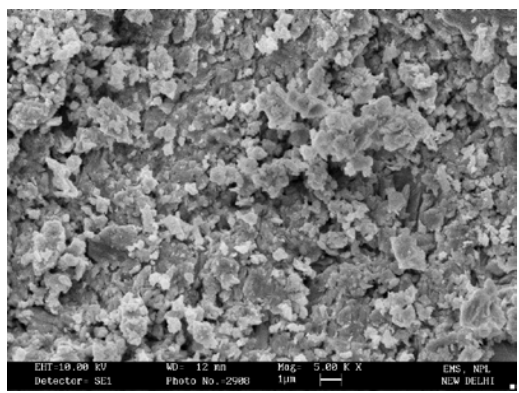

Figure 2b

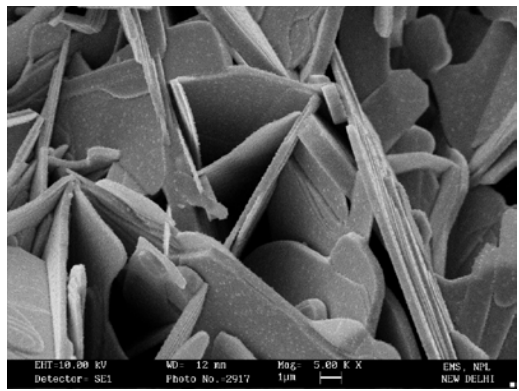

Figure 2c

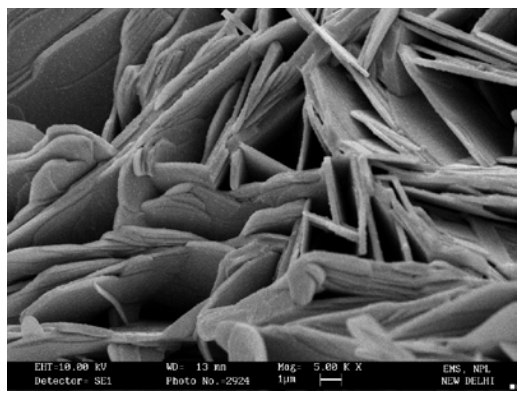

Figure 2d

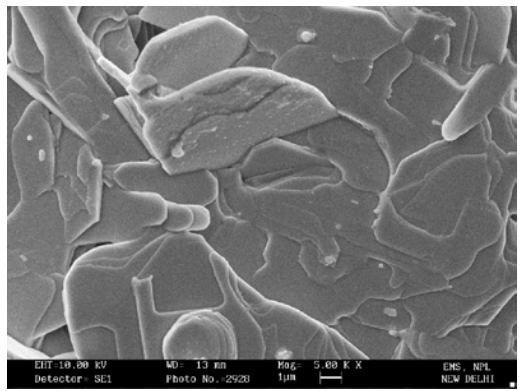

Figure 3a

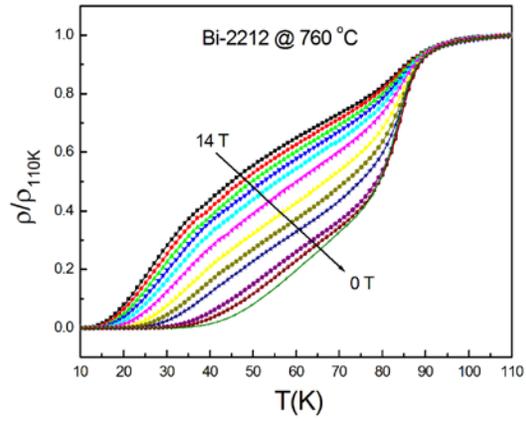

Figure 3b

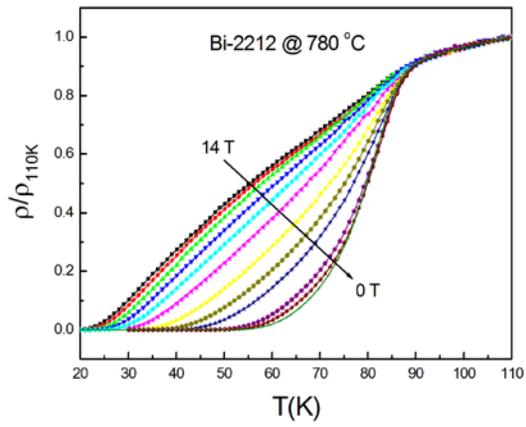

Figure 3c

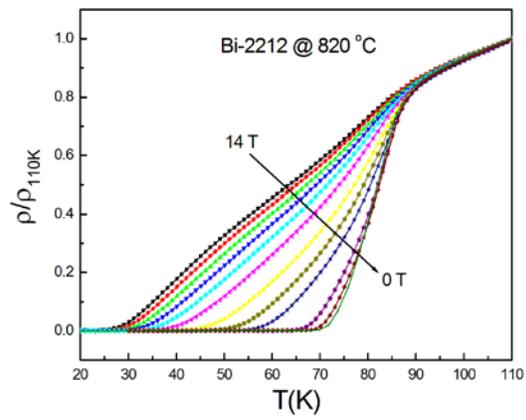

Figure 3d

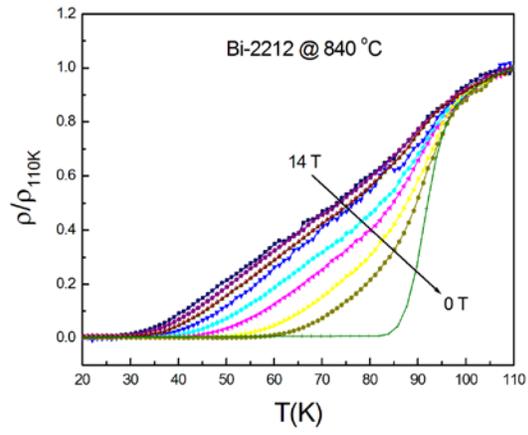

Figure 4a

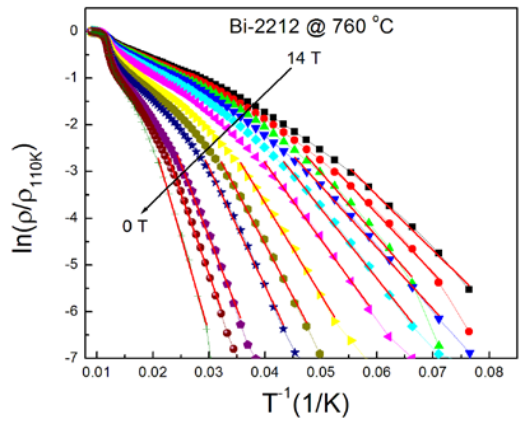

Figure 4b

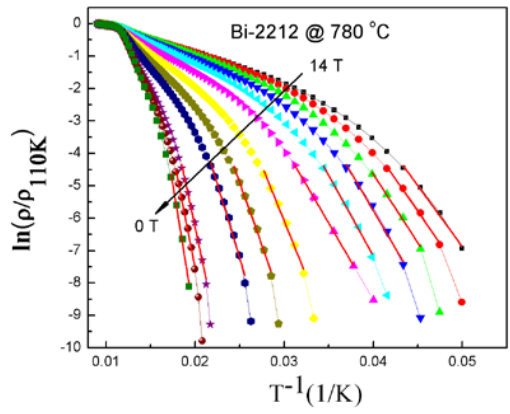

Figure 4c

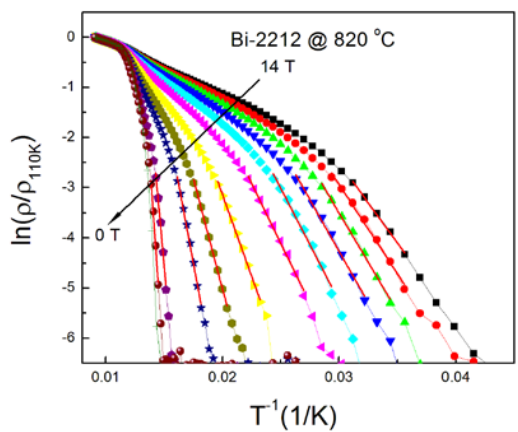

Figure 4d

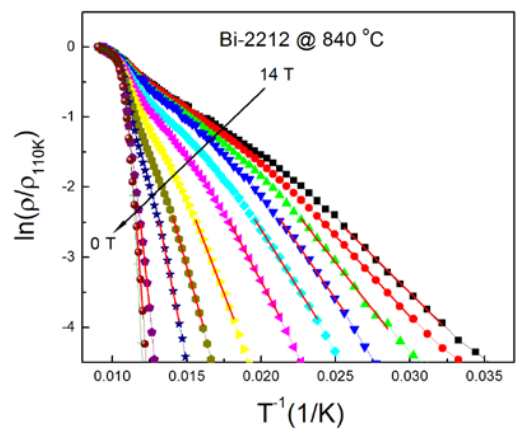

Figure 5

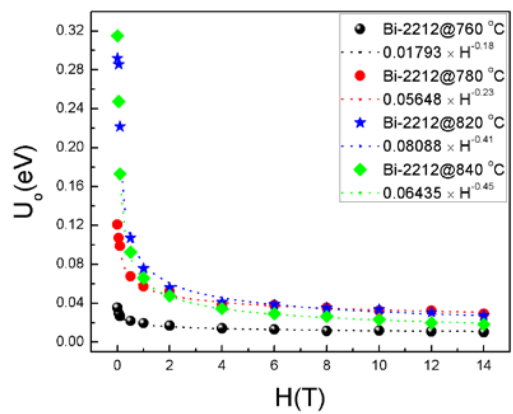

Figure 6a

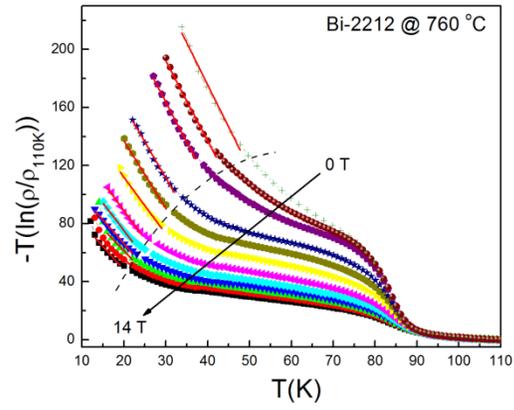

Figure 6b

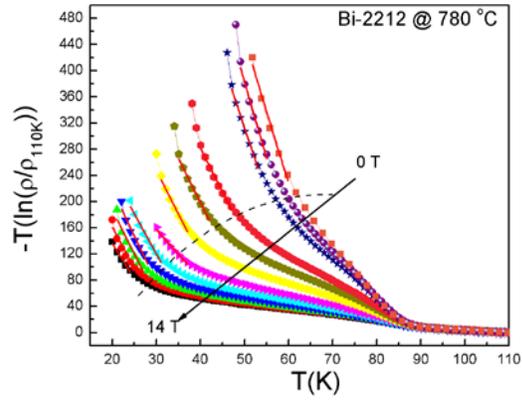

Figure 6c

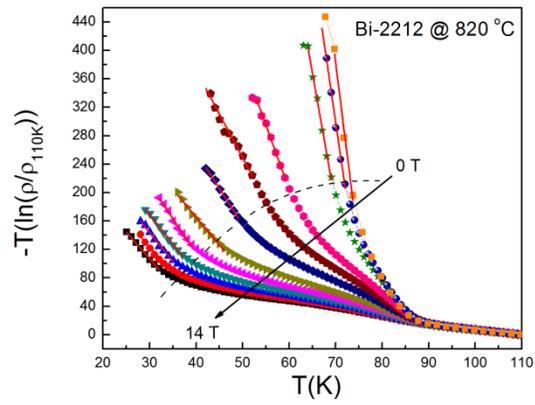

Figure 6d

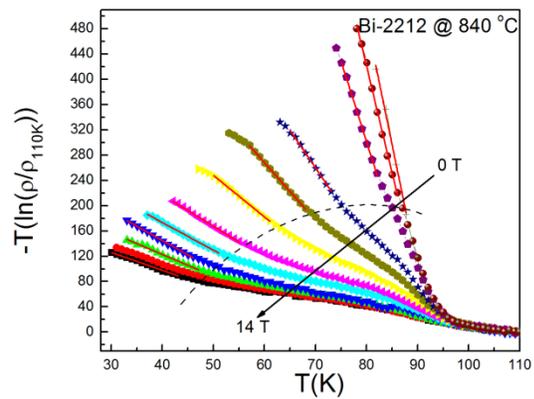